\documentclass{iopjournal}
\usepackage{cite}
\usepackage{amsmath}
\usepackage{amssymb}
\usepackage{bm}
\begin{document}

\articletype{Article type} 

\title{Interplay of field-induced magnetic phases and  magneto-transport properties in the kagome antiferromagnet $\mathrm{Tb}\mathrm{Ti}_3$$\mathrm{Bi}_4$}

\author{Sneh$^1$\orcid{0009-0001-3403-1310}, Lipika$^1$\orcid{0009-0007-1043-6173}, Subhadeep Bej$^1$\orcid{0009-0007-7832-2850}, Shobha Singh$^1$\orcid{0009-0005-1537-1482}, Ralf Koban$^2$, Walter Schnelle$^2$\orcid{0009-0007-0136-7268} and Kaustuv Manna$^{1,*}$\orcid{0000-0002-9442-2321}}

\affil{$^1$Department of Physics, Indian Institute of Technology Delhi, Hauz Khas, New Delhi 110016, India}

\affil{$^2$Max-Planck-Institute for Chemical Physics of Solids, N\"{o}thnitzer Straße 40 01187 Dresden, Germany}

\affil{$^*$Author to whom any correspondence should be addressed.}

\email{kaustuvmanna@physics.iitd.ac.in}


\begin{abstract}
Magnetic kagome materials provide a fertile platform for exploring the interplay between frustrated magnetism and electronic transport. This study investigates the anisotropic magnetic and magneto-transport properties of Tb$\mathrm{Ti}_3$$\mathrm{Bi}_4$ single crystals using \textit{dc} magnetization, \textit{ac} susceptibility, specific heat, and electrical transport measurements. For a magnetic field applied along the [100] direction, Tb$\mathrm{Ti}_3$$\mathrm{Bi}_4$ exhibits sharp metamagnetic transitions (MMTs) and fractional magnetization plateaus in the field-dependent \textit{dc} magnetization \textit{M}(\textit{H}) data, giving rise to rich magnetic phases driven by strong magnetic anisotropy and competing magnetic interactions. Here, we show that the two metamagnetic transitions in Tb$\mathrm{Ti}_3$$\mathrm{Bi}_4$ are fundamentally distinct, based on the \textit{ac} susceptibility measurements. While both transitions appear as sharp steps, only the second metamagnetic transition exhibits frequency dependence in \textit{ac} susceptibility, indicating slow non-equilibrium spin dynamics that are absent at the first metamagnetic transition. This is corroborated by the Hall resistivity, which shows a pronounced peak near the second metamagnetic transition. Analysis of the Hall response suggests that this feature may originate from an enhanced anomalous Hall contribution arising from domain-wall scattering or other non-equilibrium magnetic textures. These results demonstrate a strong coupling between field-induced magnetic phases and electronic transport in Tb$\mathrm{Ti}_3$$\mathrm{Bi}_4$ and provide new insights into the magneto-transport response of anisotropic kagome magnets.
Our results establish Tb$\mathrm{Ti}_3$$\mathrm{Bi}_4$ as an anisotropic magnetic system in which nominally similar metamagnetic steps can have distinct character, detectable through the coupling between magnetic and transport properties.
\end{abstract}

\section{Introduction}
Rare-earth-based intermetallics are an excellent class of materials for exploring complex magnetic phenomena arising from the interplay between localized 4f moments, crystalline electric field (CEF) effects, and conduction electron-mediated exchange interactions \cite{Jensen1991}. In systems where anisotropic rare-earth moments are arranged in low-dimensional or quasi-one-dimensional motifs, competing magnetic interactions can give rise to rich magnetic phase diagrams, including metamagnetic transitions, field-induced phases, and fractional magnetization plateaus reflecting nontrivial spin arrangements and magnetic frustration \cite{HeidrichMeisner2007,Siemensmeyer2008,Okunishi2003a,Aasland1997}.\\
The recently discovered RE$\mathrm{Ti}_3$$\mathrm{Bi}_4$ (RE = rare-earth atom) family of compounds has emerged as a highly tunable platform in this context, where the magnetic behavior can be systematically modified by changing the rare-earth ion \cite{Chen2024,Ortiz2023}. In these compounds, the RE atoms form quasi-one-dimensional zigzag chains along the crystallographic [100] direction, while the Ti atoms form a kagome lattice that largely governs the electronic transport. 
The incorporation of magnetic rare-earth atoms into the Ti-based kagome lattice in RE$\mathrm{Ti}_3$$\mathrm{Bi}_4$ compounds leads to diverse magnetic and electronic phenomena, such as charge-density-wave order, spin-density wave, and multiple van Hove singularities, flat bands \cite{Han2026,Park2025,Mondal2025}.\\
Within this family of compounds, Tb$\mathrm{Ti}_3$$\mathrm{Bi}_4$  exhibits highly anisotropic and complex magnetic behaviour, including fractional magnetization plateaus and multiple field-induced metamagnetic transitions for fields applied along the zigzag-chain direction of the Tb atoms. 
Previous studies have also reported an unusual coexistence of elliptical-spiral magnetic order and spin-density wave order, accompanied by band folding and a very large anomalous Hall response, highlighting the strong coupling between magnetism and electronic transport in this compound \cite{Guo2024,Ortiz2024,Cheng2026,Kushnirenko2025,Zhang2025}.\\ 

While these studies have mapped out the static magnetic phase diagram, the nature of the different metamagnetic transitions remains largely unexplored.
In this work, we present a study of the anisotropic magneto-transport properties of single-crystalline Tb$\mathrm{Ti}_3$$\mathrm{Bi}_4$ using \textit{dc} magnetization, \textit{ac} susceptibility, and electrical transport measurements, with a focus on the dynamical response across its field-induced transitions. While our \textit{dc} magnetization results reproduce the previously reported sequence of MMTs \cite{Guo2024,Ortiz2024,Cheng2026,Zhang2025}, our \textit{ac} susceptibility measurements uncover a striking contrast between them. The first transition is essentially frequency-independent, whereas the second transition exhibits a pronounced frequency dependence. Complementary magneto-transport measurements show anomalies correlated with the magnetic phase boundaries, and analysis of the Hall resistivity suggests that the peak-like features near the metamagnetic transitions likely originate from a scattering-enhanced anomalous Hall effect. In contrast to \textit{H} $\parallel$ [100], the magnetoresistance is positive and non-saturating for \textit{H} $\parallel$ [010] and [001]. For \textit{H} $\parallel$ [010], conventional Kohler scaling fails, whereas an effective temperature-dependent scaling factor improves the data collapse. These measurements complement the Hall results and characterize the anisotropic magneto-transport properties of Tb$\mathrm{Ti}_3$$\mathrm{Bi}_4$. Our results demonstrate that Tb$\mathrm{Ti}_3$$\mathrm{Bi}_4$ is a system in which seemingly comparable metamagnetic steps possess different characteristics, identifiable through the interplay of magnetic and transport measurements, and highlight the importance of incorporating dynamical probes in the study of metamagnetic systems.

\section{Experimental Details}
The single crystals of Tb$\mathrm{Ti}_3$$\mathrm{Bi}_4$ were synthesized utilizing the  Bismuth self-flux method. High-purity Tb metal pieces (99.9\%), Ti powder (99.5\%) and Bi chunks were taken in a molar ratio of 1.2:3:20 and placed in an alumina crucible. The crucible was subsequently sealed in a quartz ampoule under a vacuum pressure of about $10^{-3}$ mbar. The sealed ampule was heated up to $1050^{\circ}\text{C}$ at a controlled rate of $85^{\circ}\text{C}$/h and maintained at that temperature for 24h. Thereafter, the sample was cooled to $400^{\circ}\text{C}$ at a rate of $2^{\circ}\text{C}$/h followed by centrifugation at this temperature to remove the excess Bi flux. This yielded black, shiny hexagonal plate-like single crystals of Tb$\mathrm{Ti}_3$$\mathrm{Bi}_4$. Single crystal X-ray diffraction data were collected in gonio mode at room temperature using a PANalytical Empyrean X-ray diffractometer using a Cu K$\alpha$ ($\lambda$=1.5406 \r{A}) monochromatic source. The orientation of the as-grown single crystal was confirmed through Laue diffraction employing a PROTO Laue COS X-ray diffraction crystal orientation system. The elemental composition of the crystals was ascertained via energy dispersive X-ray spectroscopy performed on a TESCAN Magna LMU, equipped with an EDAX AMETEK Octane Elite Super EDS system. The \textit{dc} magnetization and \textit{ac} susceptibility measurements were carried out using a 7T Magnetic Property Measurement System(Quantum Design,  MPMS 3). Electrical transport and specific heat measurements were performed using a 9T Physical Property Measurement System (Quantum Design). For resistivity measurements, the crystals were shaped into a rectangular bar geometry, and measurements were executed in a standard linear four-probe geometry. The electrical contacts were established using silver paint and 25-micron platinum wire.

\section{Results and Discussion}

\subsection{Crystal Structure and magnetic ordering}

\begin{figure*}[t]
    \centering
    \includegraphics[width=1\linewidth]{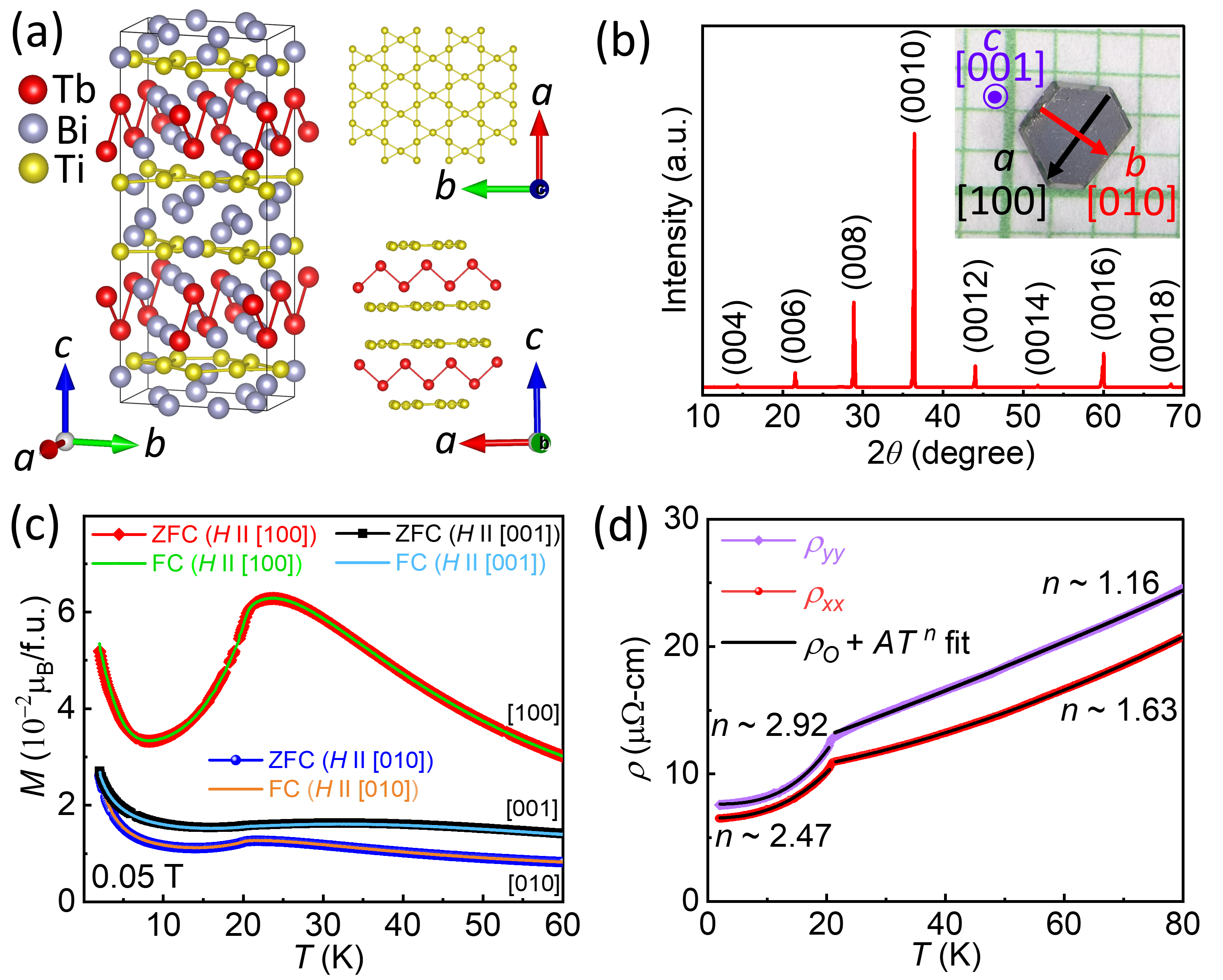}
    \caption{(a) Side view of the crystal structure of Tb$\mathrm{Ti}_3$$\mathrm{Bi}_4$. The top right panel shows the projection of the kagome network of Ti atoms on \textit{ab} plane and the bottom right panel shows the Tb atoms arranged in a zigzag chain along the \textit{a} direction. (b) XRD pattern of the Tb$\mathrm{Ti}_3$$\mathrm{Bi}_4$ single crystal, showing reflections along (00l) planes. The inset shows an as-grown Tb$\mathrm{Ti}_3$$\mathrm{Bi}_4$ single crystal along with the relevant crystallographic directions. (c) Temperature-dependent \textit{dc} magnetization \textit{M}(\textit{T}) in zero-field-cooling (ZFC) and field-cooling (FC) modes at an applied field of 0.05 T for field parallel to the [100], [010] and [001]-directions. (d) The variation of longitudinal resistivity with temperature at 0 applied field for current applied along the [010] direction, $\rho_{yy}$ and the [100] direction, $\rho_{xx}$ along with the fitting.}
    \label{fig1}
\end{figure*}

Tb$\mathrm{Ti}_3$$\mathrm{Bi}_4$ crystallises in an orthorhombic structure having space group \textit{Fmmm} (no.69). As illustrated in Fig.~\ref{fig1}~(a), the crystal structure comprises Ti atoms forming a two-dimensional kagome network in the \textit{ab}-plane and Tb atoms arranged in quasi-one-dimensional zigzag chains along the \textit{a}-direction. The Ti kagome layers are separated by alternating Bi and Tb-Bi layers. Figure~\ref{fig1}~(b) presents the X-ray diffraction pattern of a Tb$\mathrm{Ti}_3$$\mathrm{Bi}_4$ single crystal. The exclusive presence of only (00l) peaks in the diffraction pattern confirms the single crystal nature of the sample and indicates that the [001] direction is perpendicular to the flat surface of the crystal.\\ 

In order to understand the ground state physical properties of the system, we performed temperature-dependent \textit{dc} magnetization and longitudinal resistivity measurements. Figure~\ref{fig1}~(c) illustrates the temperature-dependent magnetization \textit{M}(\textit{T}) during the zero-field-cooling (ZFC) and field-cooling (FC) processes along the [100], [010] and [001] directions at an applied field of 0.05 T. No significant divergence is observed between the ZFC and FC curves along any direction. A change in slope of magnetization is discernible around 20.4 K ($T_N$), corresponding to an antiferromagnetic transition across all axes. The weak upturn at temperatures below 7 K suggests the onset of a different magnetic ordering, which is also observed in all three directions.\\
The onset of long range magnetic ordering is further corroborated by the zero field longitudinal resistivity $\rho$(\textit{T}) data as displayed in Fig.~\ref{fig1}~(d). For a current applied along both the [100] and [010] directions, resistivity increases with increasing temperature, indicative of metallic behaviour. An abrupt change of slope is observed around 20 K due to the suppression of spin-disorder scattering below the antiferromagnetic transition, with the transition temperature $T_N$ being consistent with the magnetization measurements. The resistivity anisotropy ratio between the [100] and [010] axis at 2 K is $\rho_{[100]}$(2K)/$\rho_{[010]}$(2K) $\approx$ 0.86, indicating that the electron scattering is anisotropic and depends on the orientation of magnetic moments with respect to the direction of current flow. The longitudinal resistivity for Tb$\mathrm{Ti}_3$$\mathrm{Bi}_4$ can be well fitted with the power law $\rho$ = $\rho_\circ$ + \textit{A}$T^n$, where $\rho_\circ$ is the residual resistivity, \textit{A} is the quasiparticle scattering cross section, \textit{n} is the fitting parameter. The exponent \textit{n} indicates the dominant scattering mechanism within the material, and the value of \textit{n} obtained from fitting for $\rho_{xx}$ and $\rho_{yy}$ is shown in Fig.~\ref{fig1}~(d) for both \textit{T}$<$$T_N$ and \textit{T}$>$$T_N$.
Typically, \textit{n} assumes a value of 3/2 for diffusive electron motion resulting from strong electron correlations \cite{Pfleiderer2001}, 2 for Fermi-liquid behaviour \cite{Baber1937,Abrikosov1958,VanDerMarel2011}, 3 for electron-magnon scattering \cite{Mott1964,Wilson1938,Jiang2015}, and 5 due to electron-phonon coupling \cite{Luo2015}.
The observation of $n > 2$ below $T_N$ in both cases indicates additional scattering from antiferromagnetic magnons present below $T_N$, resulting in $n > 2$ \cite{Zhao2016,Knebel2001}. In contrast, at elevated temperatures (\textit{T}$>$$T_N$), the resistivity exhibits a quasi- linear relationship with temperature, indicative of dominant electron–phonon scattering,  although deviations from strict linearity may reflect additional scattering channels.\cite{Chambers2012,Dugdale1979}\\

The effect of magnetic field on the antiferromagnetic ground state of Tb$\mathrm{Ti}_3$$\mathrm{Bi}_4$ was studied using the isothermal magnetization \textit{M}(\textit{H}) data recorded along different directions.
\begin{figure*}[t]
    \centering
    \includegraphics[width=1\linewidth]{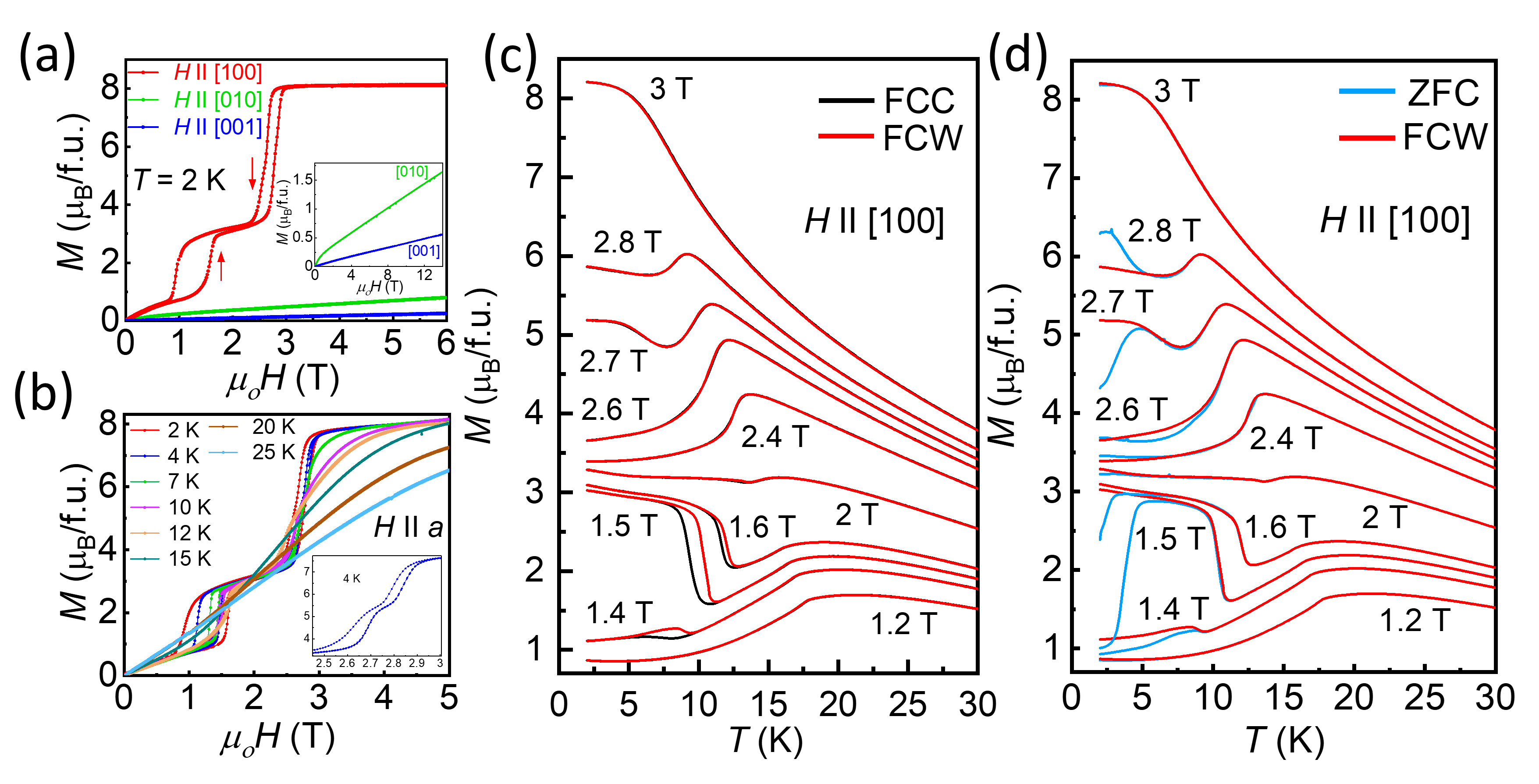}
    \caption{(a) Field-dependent isothermal \textit{dc} magnetization \textit{M}(\textit{H}) curves  at 2 K for the [100], [010] and [001]-directions. The inset shows the zoomed-in \textit{M}(\textit{H}) curves for the [010] and [001] directions measured for magnetic field up to 14 T. (b) Isothermal \textit{M}(\textit{H}) curves at different temperatures in the range of 2 to 25 K for a field along the [100] direction. Temperature-dependent variation of \textit{dc} magnetization \textit{M}(\textit{T}) at different \textit{dc} magnetic fields applied along the [100] direction for (c) field-cooled cooling (FCC) and field-cooled warming (FCW) modes and (d) zero-field cooling (ZFC) and field-cooled warming (FCW) modes.}
    \label{fig2}
\end{figure*}
Figure~\ref{fig2}~(a) displays the \textit{M}(\textit{H}) curves at 2 K for field parallel to the [100], [010] and [001] axes. For $H \parallel [010]$ and $H \parallel [001]$, the magnetization increases almost linearly with the field, as expected for an antiferromagnet, and remains unsaturated up to 14 T. In contrast, for $H \parallel [100]$, the \textit{M}(\textit{H}) curve exhibits two pronounced step-like metamagnetic transitions (MMTs) at about 1.5 T and 2.6 T. The first transition leads to a plateau at nearly 1/3 of the saturation magnetization, while the second transition drives the system into a saturated state above 3 T. Narrow hysteresis loops between the up- and down-sweep curves indicate the first-order nature of these field-induced transitions. A weaker anomaly near 2/3 of the saturation magnetization is also resolved in between the second MMT, and is more clearly visible at 4 K in the inset of Figure~\ref{fig2}~(b). These MMTs and fractional plateaus likely originate from field-induced spin reorientation, as reported in many other chain compounds \cite{Mokhtari2025,Kageyama1997,Jiang2026,Ma2023}. These results agree with earlier reports \cite{Guo2024,Ortiz2024,Cheng2026,Kushnirenko2025,Zhang2025} and confirm the easy axis anisotropy along the [100] axis. Previous reports attributed the first MMT to flipping of one-third of the moments in a layer, giving rise to the 1/3 plateau, while the 2/3 kink was similarly associated with reversal of two-thirds of the moments. As evident in Figure~\ref{fig2}~(b), the hysteresis weakens with increasing temperature and above 20 K the transitions disappear, and magnetization increases almost linearly with field.\\

To further explore the field evolution of magnetic phases for $H \parallel [100]$, temperature-dependent magnetization measurements \textit{M}(\textit{T}) were carried out under different \textit{dc} fields using field-cooled cooling (FCC), field-cooled warming (FCW) and zero-field-cooled (ZFC) protocols, as shown in Fig.~\ref{fig2}~(c) and  Fig.~\ref{fig2}~(d). The cusp at $T_N$ shifts to lower temperatures with increasing field, as expected for antiferromagnets. As the field increases, the additional upturn observed at lower temperatures corresponds to the first MMT, marking the transition from the AFM state to the 1/3 plateau state. The thermal hysteresis between FCC and FCW across the AFM to 1/3 transition supports its first-order nature. From 2 T to 2.6 T the system transforms directly from the paramagnetic state (PM) to the 1/3 phase. At 2.7 T, the cusp around 11 K corresponds to the PM to 1/3 transition, while the upturn near 9 K reflects the second MMT from the 1/3 to the 2/3 state. At 2.8 T, the cusp near 9 K, indicates a direct transition from the PM state to the 2/3 state. Above this field, no further anomalies are observed and the magnetization saturates at low temperature. Figure~\ref{fig2}~(d) shows that the ZFC and FCW curves bifurcate at low temperatures for fields above 1.4 T. In magnetically ordered materials, this may arise due to magnetocrystalline anisotropy, anisotropy resulting from reduced dimensions, competing magnetic interactions, lattice deformation and phase coexistence etc \cite{Kumar2021,Roy2004,Chen2007,Levin2002,LampenKelley2020,Joy1998,Kumar1998}. The origin of this bifurcation is explained in more detail in Fig S3.(c) and Fig S3.(d) in the SM. 

\subsection{Field- and temperature- dependent \textit{ac} susceptibility}

\begin{figure}
    \centering
    \includegraphics[width=1\linewidth]{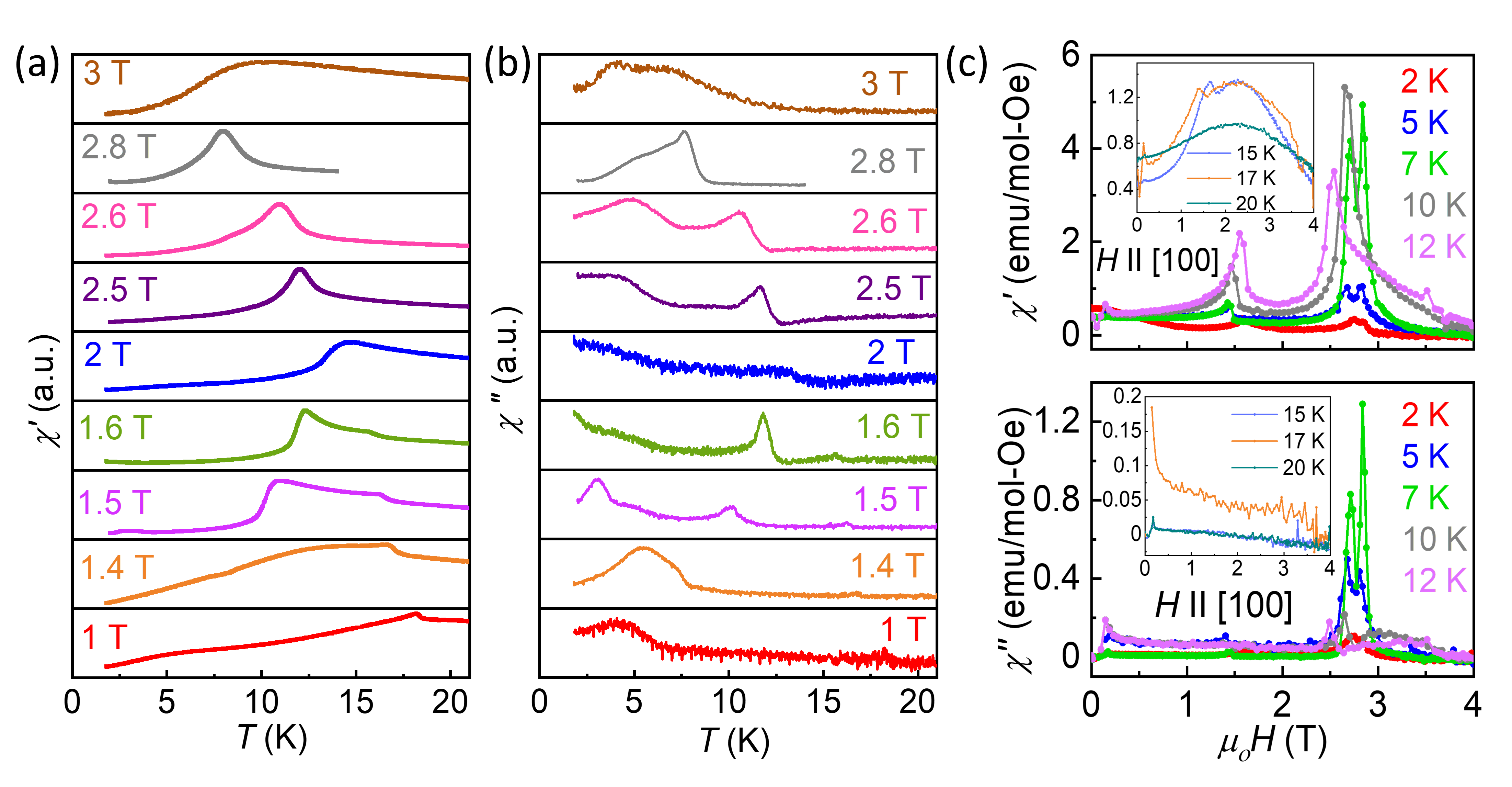}
    \caption{Temperature-dependent (a) real ($\chi^{\prime}$(T)) and (b) imaginary ($\chi^{\prime\prime}$(T)) parts of the \textit{ac} susceptibility with an \textit{ac} excitation field of 9 Oe and excitation frequency of 57 Hz superimposed over different \textit{dc} fields for $H \parallel [100]$. (c) Field-dependent $\chi^{\prime}$ and  $\chi^{\prime\prime}$ with an  \textit{ac} excitation field of 9 Oe and excitation frequency of 57 Hz superimposed over a \textit{dc} field at different temperatures for $H \parallel [100]$. The inset shows the $\chi^{\prime}$ vs field and $\chi^{\prime\prime}$ vs field curves for temperatures above 15 K.}
    \label{fig3}
\end{figure}

We performed field- and temperature-dependent \textit{ac} susceptibility measurements with an \textit{ac} excitation field of 9 Oe and excitation frequency of 57 Hz in order to gain insights into the nature of magnetic transitions. Figure ~\ref{fig3}~(a) and Fig.~\ref{fig3}~(b) displays the variation of the real ($\chi^{\prime}$) and the imaginary ($\chi^{\prime\prime}$) part of the \textit{ac} susceptibility with temperature respectively, having different \textit{dc }fields superimposed on the \textit{ac} driving field for $H \parallel [100]$. The $\chi^{\prime}$(\textit{T}) curves are consistent with the \textit{dc} magnetization data in Fig.~\ref{fig2}. At 1 T, a weak cusp is observed, corresponding to the PM-to-AFM transition seen in the \textit{dc M}(\textit{T}) curve. With increasing field from 1 T to 1.6 T, this cusp shifts to lower temperature, while the corresponding $\chi^{\prime\prime}$(\textit{T}) anomaly remains weak, as expected for an antiferromagnetic transition. At 1.5 T and 1.6 T, an additional peak appears near 10 K, associated with the transition from the low-field AFM to the $1/3$ magnetization plateau phase upon cooling. This transition is accompanied by a pronounced peak in $\chi^{\prime\prime}$(\textit{T}), indicating finite energy dissipation. In long-range ordered magnetic systems, such losses commonly arise from irreversible processes such as domain-wall motion, pinning and depinning, domain reconstruction, or irreversible rotation of magnetization \cite{Balanda2013,Chen1996}.\\
At 2 T and 2.5 T, the peaks in $\chi^{\prime}$(\textit{T}) correspond to the direct PM to $1/3$ transition, whereas the feature at 2.8 T is assigned to the PM to $2/3$ transition, consistent with the \textit{dc M}(\textit{T}) data. Both transitions are accompanied by finite jumps in $\chi^{\prime\prime}$(\textit{T}), again suggesting the involvement of domain-related dissipative processes. In the fully field-polarized state at 3 T, only a broad hump remains in $\chi^{\prime}$(\textit{T}). Field-dependent \textit{ac} susceptibility measurements, shown in Fig.~\ref{fig3}~(c), further confirm the field-induced phase transitions observed in the \textit{dc M}(\textit{H}) data in Fig.~\ref{fig2}. Peaks in $\chi^{\prime}$(\textit{H}) appear at the transition fields, the peak near 1.5 T corresponds to the AFM to $1/3$ transition, while the double-peak structure between 2.5 T and 3 T reflects successive $1/3$ to $2/3$ and $2/3$ to field-polarized phase transitions. These double peaks occur only over a limited temperature range, consistent with the \textit{dc} magnetization results. The simultaneous non-zero $\chi^{\prime\prime}$(\textit{H}) signal at these metamagnetic transitions further indicates the energy losses associated with these transitions.

\subsection{Frequency-dependent \textit{ac} susceptibility}
\begin{figure}
    \centering
    \includegraphics[width=1\linewidth]{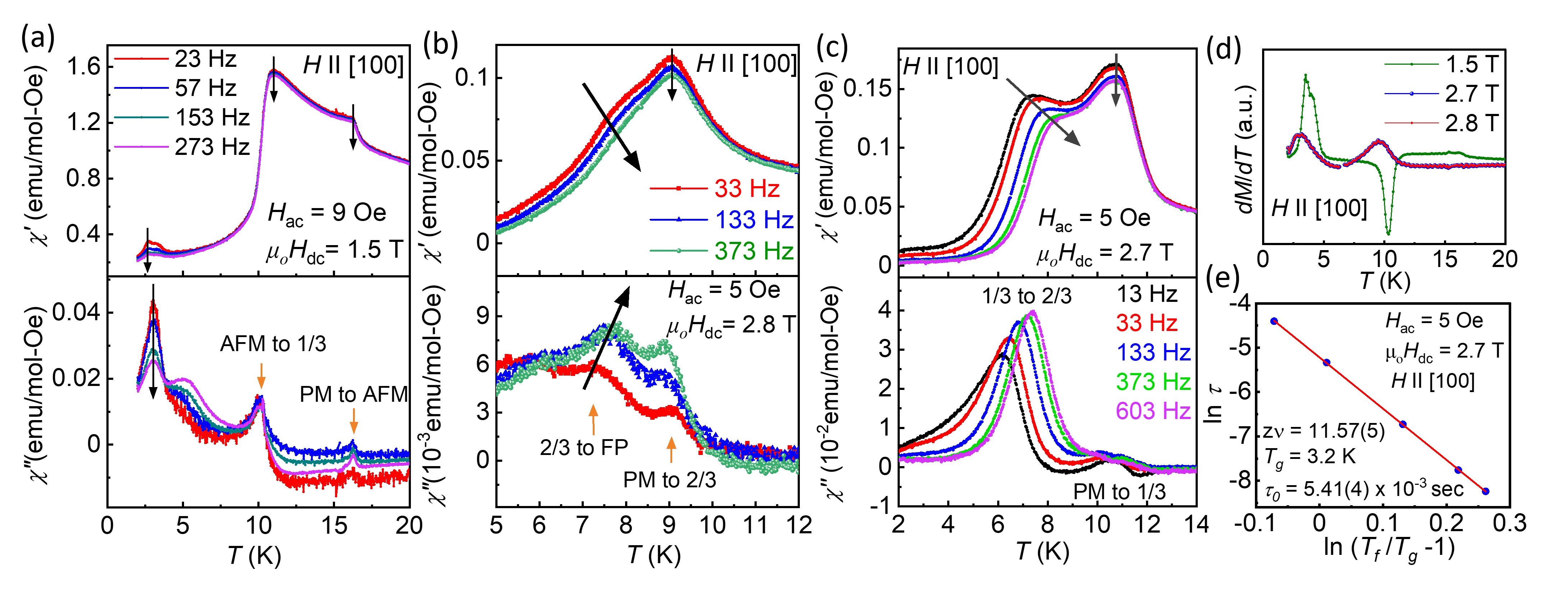}
    \caption{Temperature-dependent real ($\chi^{\prime}$(\textit{T}))and imaginary ($\chi^{\prime\prime}$(\textit{T})) parts of the \textit{ac} susceptibility at different excitation frequencies and (a) an \textit{ac} driving field amplitude of 9 Oe, superimposed over a \textit{dc} field of 1.5 T, (b) an \textit{ac} driving field amplitude of 5 Oe, superimposed over a \textit{dc} field of 2.8 T and (c) 2.7 T for $H \parallel [100]$. (d) Temperature-dependent $dM/dT$ at \textit{dc} fields of 1.5T, 2.7 T and 2.8 T applied along [100] direction, as derived from ZFC- M(\textit{T}) data. (e) The best fit of ln $\tau$ vs ln ($T_f$/$T_g$-1) corresponding to $H_{dc}$ = 2.7 T along with the fitted parameters.}
    \label{fig4}
\end{figure}
To understand the spin-relaxation dynamics associated with different transitions, frequency-dependent \textit{ac} susceptibility measurements were performed, as shown in Fig.~\ref{fig4}. Under finite \textit{dc} fields, the peaks or humps associated with the PM to AFM and AFM to $1/3$ transitions [Fig.~\ref{fig4}~(a)], the PM to $2/3$ transition [Fig.~\ref{fig4}~(b)] and the PM to $1/3$ transition [Fig.~\ref{fig4}~(c)] are nearly frequency independent. Notably, all transitions with a superposed \textit{dc} field of 1.5 T exhibit no frequency-dependent peak shift, as illustrated in Fig.~\ref{fig4}~(a). However, the lower temperature peak in the second MMT region demonstrates a clear frequency-dependent peak shift, as shown for an applied \textit{dc} field of 2.8 T and 2.7 T in Fig.~\ref{fig4}~(b) and ~\ref{fig4}~(c) respectively. This shift might occur due to slow spin dynamics arising from competing AFM–FM interactions, coexistence of competing magnetic phases, presence of some spin textures or domain dynamics in this field region \cite{Mulder1981,Levatic2014,Kumar2025, Tsurkan2001}.\\
 The transitions observed in the temperature-dependent \textit{ac} susceptibility data are also evident in the temperature-dependent $dM/dT$ data derived from ZFC-\textit{M}(\textit{T}) data, as shown in Fig.~\ref{fig4}~(d). At an applied \textit{dc} field of 2.7 T, as displayed in Fig.~\ref{fig4}~(c), for the peak corresponding to the transition from the $1/3$ to the $2/3$ phase, a substantial frequency-dependent peak shift is observed. This shift was analyzed using the critical slowing down model \cite{Mydosh2014,Adeniji2025,Manna2015}, which is given by, $\tau = \tau_0 (T_f/T_g-1)^{-z\nu}$, where $\tau$= 1/2$\pi$\textit{f} is the relaxation time corresponding to the measurement frequency \textit{f}, $\tau_0$ is the microscopic spin flipping time, $z\nu$ is the dynamical critical exponent, $T_f$ is the spin freezing temperature at frequency \textit{f} and $T_g$ is the spin freezing temperature for f $\rightarrow$ 0. Corresponding to this equation, the best fit in the ln $\tau$ vs ln $(T_f/T_g-1)$ curve in Fig.~\ref{fig4}~(e) obtained by employing the method described in \cite{RoyChowdhury2023} yielded $\tau_0$ = 5.41(4) x $10^{-3}$ sec. The relatively large value of $\tau_0$ indicates slow spin relaxation dynamics across the second metamagnetic transition. Such slow relaxation may originate from domain dynamics or other non-equilibrium magnetic processes associated with slow collective spin dynamics due to the presence of correlated spin textures near this transition \cite{Madduri2020,Levatic2016,Yadav2023,Lipika2026}. Thus, the frequency-dependent \textit{ac} susceptibility measurement demonstrates that the spin dynamics and characteristics associated with the two MMTs are fundamentally distinct.

\subsection{Temperature-dependent specific heat}
\begin{figure*}[t]
    \centering
    \includegraphics[width=1\linewidth]{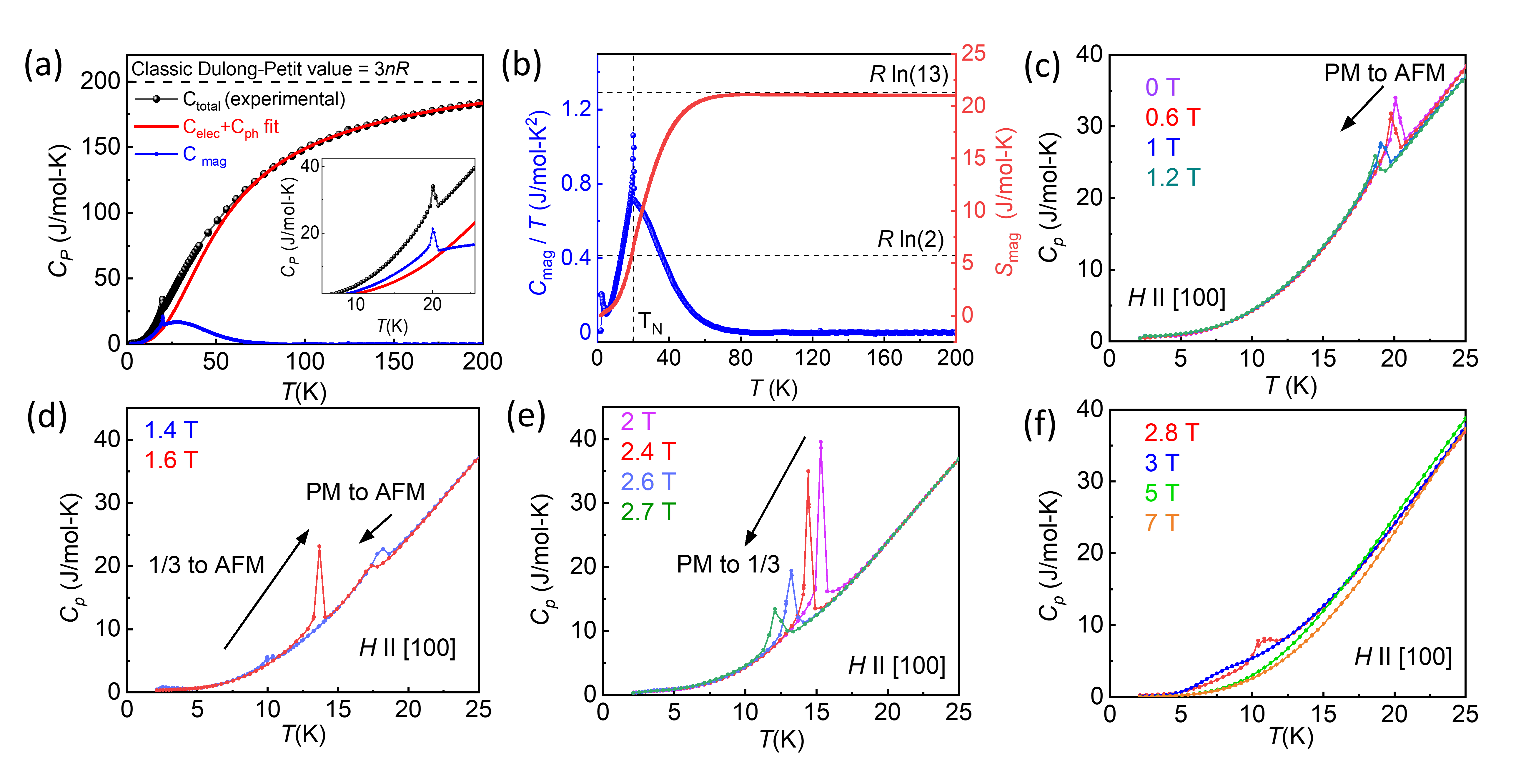}
    \caption{(a) Temperature dependence of the specific heat $C_p$(T) at zero applied field. The solid red line shows the fitting for the combined electronic $C_\mathrm{elec}$ and lattice specific heat $C_\mathrm{ph}$ contributions to the total specific heat. The blue curve shows the variation of the magnetic specific heat $C_\mathrm{mag}(T)$ with temperature. Inset shows the zoomed in curves near $T_N$. (b) Variation of  $C_\mathrm{mag}(T)$/T and the magnetic entropy $S_\mathrm{mag}$ (\textit{T}) with temperature. 
    Temperature-dependent specific heat $C_p$(\textit{T}) for different fields applied along the [100] direction, highlighting the (c) paramagnetic-antiferromagnetic transition, (d) paramagnetic-antiferromagnetic and antiferromagnetic-1/3 phase transition (e) paramagnetic-1/3 phase transition and (f) the paramagnetic-2/3 and field-polarized phase transitions.}
    \label{fig5}
\end{figure*}
The temperature dependence of the specific heat $C_p$ at zero magnetic field is shown in Fig.~\ref{fig5}~(a). Reliable data with estimated inaccuracy less than $\pm 4$\,\% could be obtained up to 200\,K. The pronounced lambda-shaped peak near 20\,K indicates the antiferromagnetic to paramagnetic phase transition.
For $T \approx 200$\,K the specific heat approaches the Dulong-Petit value of $3nR \approx 200$ J mol$^{-1}$ K$^{-1}$, where $n = 8$ is the number of atoms per formula unit and $R$ is the universal gas constant. For TbTi$_3$Bi$_4$, various contributions to the specific heat have to be considered. The measured data may be described by $C_p(T) =  C_\mathrm{ph} + C_\mathrm{elec} + C_\mathrm{mag}$, where $C_\mathrm{ph}$ is the lattice contribution attributable to phonons, $C_\mathrm{elec}$ is the electronic contribution from conduction electrons, and $C_\mathrm{mag}$ is the contribution arising due to the Tb $4f$ moments and the exchange interactions among them. $C_\mathrm{ph}$ may be modelled by a combination of lattice modes through the Debye and Einstein formalisms as $C_{ph}$ = $\eta$$C_{V,Debye}$(\textit{T})+(1-$\eta$)$C_{V,Einstein}$(\textit{T}), where $\eta$ is the weighting factor, representing the relative contribution of Debye modes to the overall lattice heat capacity. $C_{V,Debye}$(\textit{T}) and $C_{V,Einstein}$(\textit{T}) are defined as\\
\[
C_{V,Debye}(T) = 9\textit{n}\textit{R} \left( \frac{T}{\theta_D} \right)^3 \int_0^{\theta_D/T} \frac{x^4 e^x}{(e^x - 1)^2} dx
\]
and \[
C_{V,Einstein}(T) = 3\textit{n}\textit{R} \left(\frac{\theta_E}{T}\right)^2 \frac{e^{\theta_E/T}}{(e^{\theta_E/T}-1)^2}
\], where $\theta_D$ and $\theta_E$ are Debye and Einstein temperatures respectively \cite{Gopal1966,Tari2003}.
The electronic contribution is linear in $T$ and $C_\mathrm{elec} = \gamma T$, where the Sommerfeld coefficient $\gamma$, is related to the density of states at Fermi level $D(E_\textrm{F})$, for free itinerant charge carriers by the relation $\gamma$=$\pi^2$$k_{B}^{2}$\textit{D($E_F$)}/3.\\
Relevant for Tb$^{3+}$ ions and their $4f$-derived contribution to the specific heat is the $^7F_6$ multiplet of this ion. With Tb on Wyckoff position $8i$ with low $mm2$ symmetry, the multiplet is split in $2J+1 = 13$ singlets. The typical total splitting of this crystal field multiplet is 200--300\,K. In specific heat, the Schottky anomaly from thermal excitations of these states and exchange interactions of the $4f$ spins has a total entropy of $R ln{13}$. A unique analysis of the numerous contributions to the specific heat for a complex metallic ternary compound is not possible. However, to extract some salient features, we adopted the following procedure : First, the available experimental data was fitted with $C_\mathrm{ph} + C_\mathrm{elec}$ in the range 80 K-200 K. The fit curve is presented in Fig.~\ref{fig5}~(a) (with the resulting Sommerfeld parameter $\gamma \approx 30$ mJ mol$^{-1}$ K$^{-2}$). The $4f$ contribution to the specific heat is then determined by subtracting the obtained electronic and lattice contributions from the total specific heat, with the resultant $C_\mathrm{mag}$ depicted in Fig.~\ref{fig5}~(a). The magnetic entropy $S_\mathrm{mag}$(\textit{T}) was calculated from $C_\mathrm{mag}$(\textit{T}) as $S_\mathrm{mag}$(\textit{T})=$\int_{0}^{T}\frac{C_\mathrm{mag}(T)}{T}$\, d\textit{T} and the $C_\mathrm{mag}(T)$ and $S_\mathrm{mag}(T)$ curves are plotted in Fig.~\ref{fig5}~(b). Although the magnetic entropy at $T_\mathrm{N}$ is about $R ln{2}$, the short-range contributions above $T_\mathrm{N}$ are easily visible and they amount to an entropy of about $R ln{3}$, indicating that there are three states involved in the magnetic ordering.  The total magnetic entropy for the Tb$^{3+}$ ion of $R ln{13}$ is consistently recovered by our model.

To further investigate the field-induced phase transitions, the temperature-dependent specific heat \textit{$C_p$}(\textit{T}) was measured under different applied \textit{dc} magnetic fields. The peaks observed in the specific heat correspond to phase transitions between different magnetic phases and are consistent with the magnetization data. As shown in Fig.~\ref{fig5}~(c), when the field is increased from 0 to 1.2 T, the $\lambda$ peak corresponding to $T_N$ shifts towards a lower temperature, broadening and diminishing in intensity, as is typical for antiferromagnets. Above an applied field of 1.4 T in Fig.~\ref{fig5}~(d), in addition to the $\lambda$ peak, another sharp delta-like peak is observed at a lower temperature, which shifts towards higher temperatures with an increasing magnetic field. This sharp peak corresponds to the first-order transition from the AFM to the 1/3rd state and occurs at approximately the same temperature as observed in magnetization measurements. At higher magnetic fields above 2 T, the two peaks merge resulting in a single peak that corresponds to the direct PM to 1/3rd phase transition. The position of this peak shifts towards lower temperatures with increasing field, as evident in Fig.~\ref{fig5}~(e). The cusp at 2.8 T corresponds to the PM to 2/3rd phase transition and eventually smears out completely above 3 T (Fig.~\ref{fig5}~(f)), as the application of high magnetic fields induces the field polarized state to transition to a low-entropy configuration, accounting for the broadening of the $\lambda$ -type peak observed above 3 T \cite{Souza2019}.

\subsection{Electrical transport properties}
\begin{figure}
    \centering
    \includegraphics[width=1\linewidth]{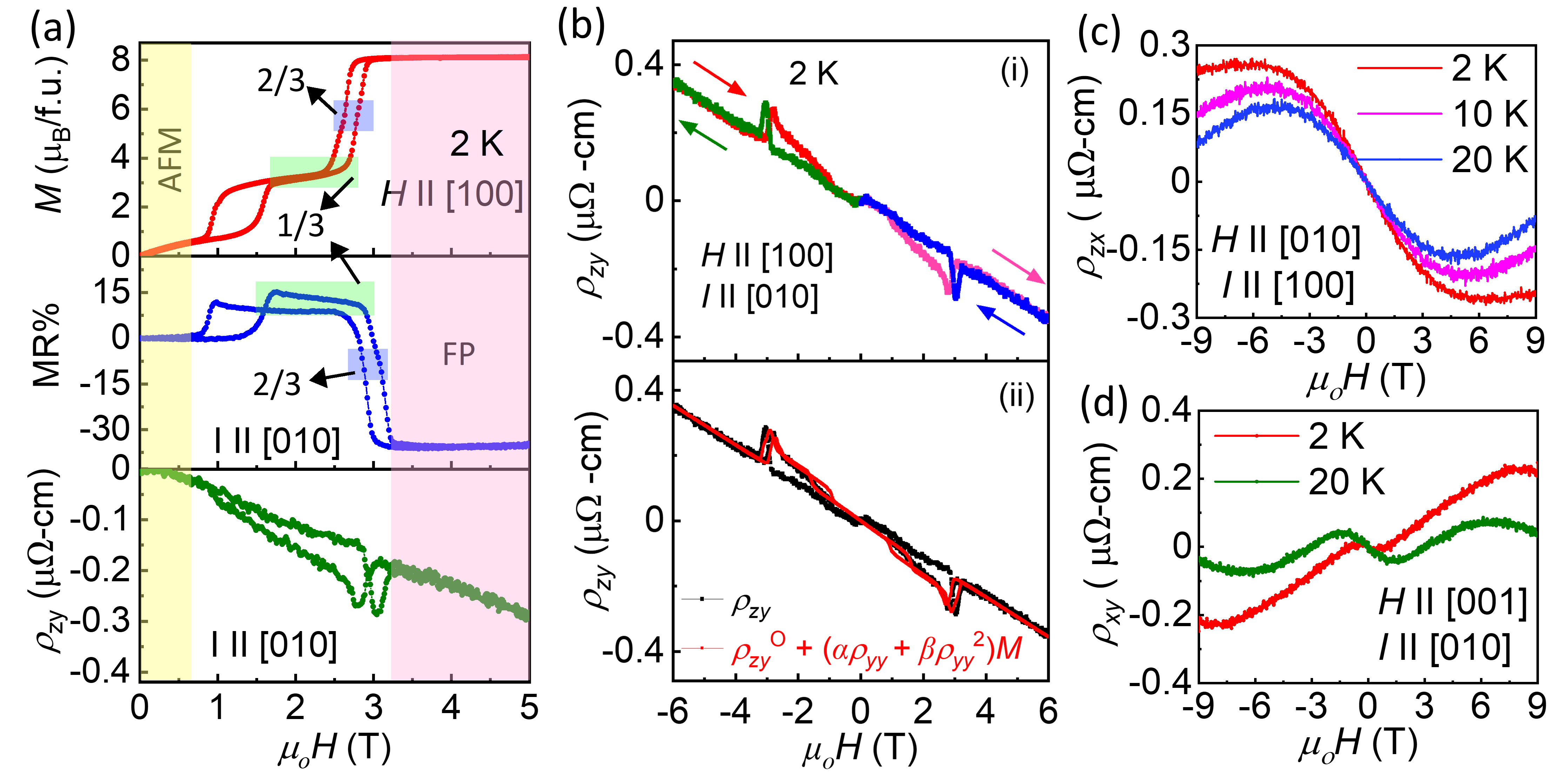}
    \caption{(a)Field dependence of magnetization, magnetoresistance, and Hall resistivity at 2 K for \textit{H} $\parallel$ [100]. Top panel shows the magnetization \textit{M}(\textit{H}), showing two field-induced metamagnetic transitions and fractional magnetization plateaus near 1/3 and 2/3 of the saturation moment. The middle panel shows the  transverse magnetoresistance (MR) for current I $\parallel$ [010], displaying clear anomalies at the metamagnetic transition fields. The bottom panel shows the Hall resistivity exhibiting distinct peak-like features near the metamagnetic transitions. The shaded regions denote the antiferromagnetic (AFM), 1/3, 2/3 and field-polarized (FP) phases. (b)(i) Field dependence of the Hall resistivity at 2 K for field \textit{H} $\parallel$ [100] and current I $\parallel$ [010]. Arrows indicate the up-sweep and down-sweep curves. (ii) The experimental Hall resistivity data along with the simulated total Hall resistivity data modelled as $\rho_{zy}$ = $\rho_{zy}^O$ + ($\alpha$$\rho_{yy}$ + $\beta$$\rho_{yy}^2$)\textit{M}. Field-dependence of the Hall resistivity at different temperatures for (c) \textit{H} $\parallel$[010], I $\parallel$ [100]  and (d) \textit{H} $\parallel$ [001], I $\parallel$ [010]. }
    \label{fig6}
\end{figure}
\begin{figure*}[t]
    \centering
    \includegraphics[width=0.9\linewidth]{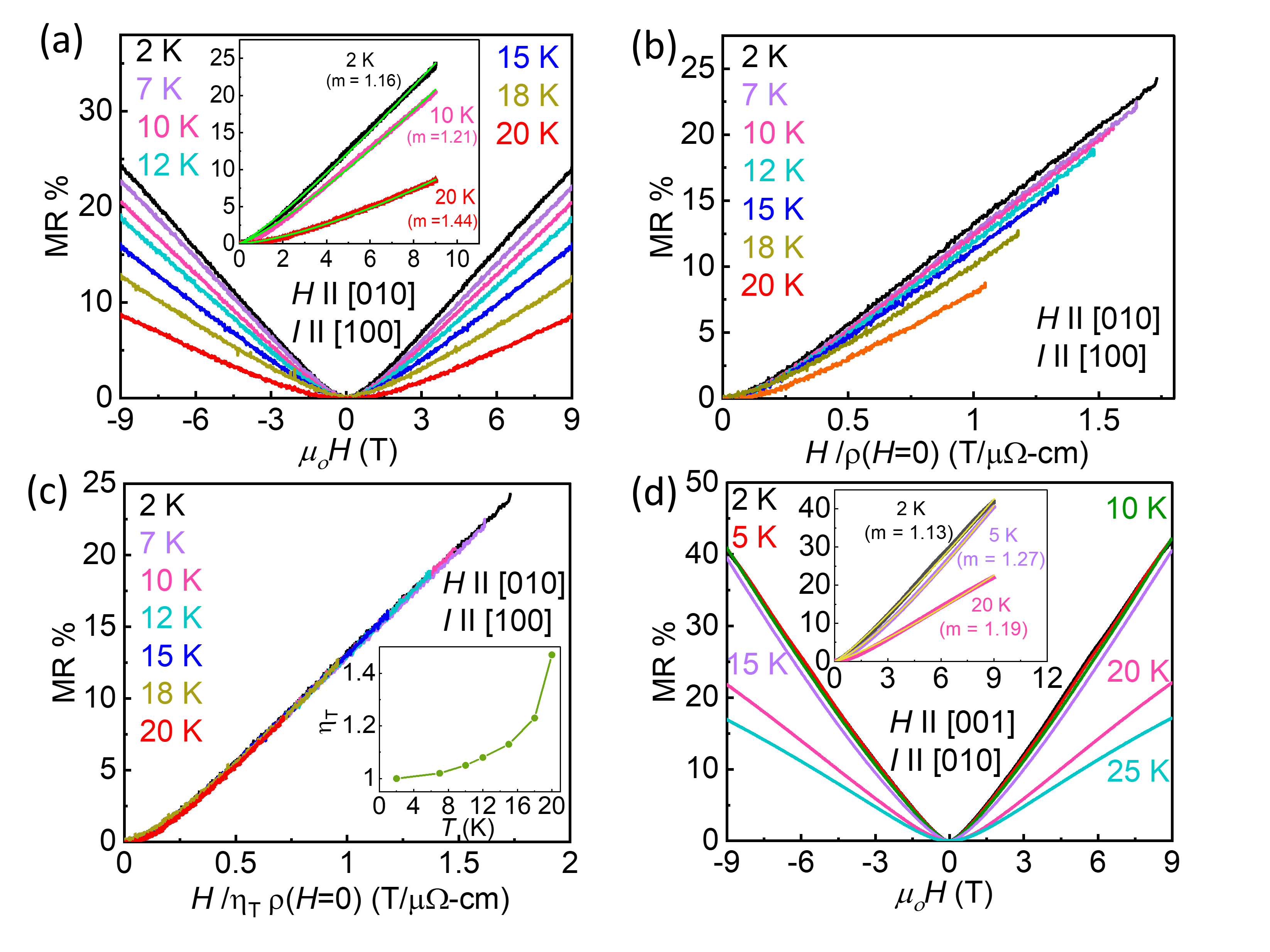}
    \caption{(a) Field-dependent MR data for field applied along the [010] direction in the temperature range 2-20 K. The inset shows the fitting of the MR vs field data at 2K, 10K and 20 K with $MR \propto H^m$. (b) The Kohler's plots and (c) the extended Kohler's plots for the MR data at different temperatures for \textit{H} $\parallel$ [010]. The inset of (c) shows the variation of the parameter $\eta_T$ with temperature. (d) Field dependence of MR at different temperatures for \textit{H} $\parallel$ [001]. The inset shows the fitting of the MR vs field data at 2K, 5K and 20 K with $MR \propto H^m$.}
    \label{fig7}
\end{figure*}
To elucidate the impact of the complex magnetic phases on charge transport in Tb$\mathrm{Ti}_3$$\mathrm{Bi}_4$, comprehensive temperature- and field-dependent resistivity measurements were performed along all three crystallographic directions. Figure ~\ref{fig6}~(a) shows the magnetoresistance (MR) at 2 K for \textit{H} $\parallel$ [100]. A clear hysteresis between field-up and field-down sweeps is evident, with distinct MR jumps coinciding with the two metamagnetic transitions seen in the \textit{M}(\textit{H}) data, highlighting a strong coupling between the magnetic phases and charge transport. In the low field antiferromagnetic region, MR is small and nearly field-independent. It shows a sudden positive jump at the first metamagnetic transition from nearly 0\%  upto 1.3 T to about 15 \%  at 1.8 T and changes only slightly on the 1/3 plateau indicating enhanced spin-disorder scattering in this phase. Near the second transition, MR shows a large negative jump from around 10.5\% at 2.8 T to - 35\% near 3 T, followed by saturation in the field-polarized state, consistent with suppressed magnetic scattering due to moment alignment. A weak feature between these regimes corresponds to the intermediate 2/3 phase. For \textit{H} $\parallel$ [100], the Hall resistivity $\rho_{zy}$ (current \textit{I} $\parallel$ [010])  also shows anomalies at the metamagnetic transitions, including a peak-like feature between the second transition and saturation.\\
Panel (i) of Fig.~\ref{fig6}~(b) illustrates that at magnetic fields exceeding 3 T, where magnetization saturates according to the \textit{M}(\textit{H}) data, the Hall resistivity demonstrates an approximately linear behavior with a negative slope. This negative slope at high fields suggests that the Hall response is electron-like within this field range. In this field-polarized regime, the ordinary contribution to the total Hall resistivity, denoted as $\rho_{zy}^O$ can be expressed as $R_0$$\mu_0$\textit{H}, where $R_0$ represents the ordinary Hall coefficient. By performing a linear fit on the high-field data, the magnitude of $R_0$ was determined to be approximately $\sim$ 6.8 x $10^{-10}$ $\Omega$-m/T. It is important to note that a linear Hall slope does not preclude contributions from multiple bands, as evidenced by the nonlinear Hall response observed for other field orientations. Consequently, the overall transport cannot be adequately described by a simple single-band model. The linear model is employed here solely to estimate an ordinary Hall background for the anomalous Hall analysis.
To elucidate the origin of this peak-like anomaly in the Hall resistivity $\rho_{zy}$, we decompose the total Hall resistivity into the ordinary $\rho_{zy}^O$ and the additional anomalous Hall $\rho_{zy}^{A}$ contributions. The anomalous Hall term $\rho_{zy}^A$, which depends on the magnetization, is modelled as $\rho_{zy}^A$ = ($\alpha$$\rho_{yy}$ + $\beta$$\rho_{yy}^2$)\textit{M} \cite{Shang2021,Liu2017}. Here the first term $\alpha\rho_{yy}$\textit{M} corresponds to the extrinsic skew scattering contribution and the second term  $\beta$$\rho_{yy}^2$\textit{M} denotes the side jump contribution as well as the intrinsic contribution associated with the Berry phase, $\rho_{yy}$ is the longitudinal resistivity and \textit{M} is the magnetization at the corresponding temperature and field values. The total Hall resistivity data simulated with these equations is presented in panel (ii) of Fig.~\ref{fig6}~(b), where the values for the coefficients are taken as $\alpha$ = 6 x $10^{-4}$ g $\mathrm{emu^{-1}}$ and $\beta$ = -120 g $\mathrm{emu^{-1}}$ $\mathrm{\Omega^{-1}}$$\mathrm{cm^{-1}}$.  
The simulated curve closely aligns with the experimental data, indicating that the peaks observed in the experimental Hall resistivity data are most likely due to the anomalous Hall effect. The position of these peaks coincides with the second metamagnetic transition, where slow spin relaxation was observed from the \textit{ac} susceptibility data. This suggests that this kind of anomalous Hall response observed for Tb$\mathrm{Ti}_3$$\mathrm{Bi}_4$ likely has a significant extrinsic, domain-wall-related or spin texture related component. Recent investigations into a similar compound, Gd$\mathrm{Ti}_3$$\mathrm{Bi}_4$, have associated these peaks with domain wall effects as observed by magnetic force microscopy (MFM), resulting in an anomalous Hall resistivity enhanced due to domain-wall skew scattering \cite{Cheng2024}.\\ 
In contrast to the \textit{H} $\parallel$ [100] configuration, when the field is applied parallel to [010] or [001] axis, no spikes or abrupt jumps are observed in the Hall resistivity data as shown in Fig.~\ref{fig6}~(c) and Fig.~\ref{fig6}~(d) respectively. The highly non-linear data even above $T_N$ suggests a complex charge-transport response, potentially involving contributions from multiple carrier types or multiband transport. Figure ~\ref{fig7}~(a) shows the field-dependent MR at different temperatures for $H \parallel [010]$ and $I \parallel [100]$. Across all temperatures, the MR remains positive and nonsaturating, with its magnitude decreasing as the temperature increases. To examine the scaling behavior, the data were initially analyzed using Kohler’s rule, MR = \textit{f}(\textit{H}/$\rho$(\textit{H}=0)) \cite{Pippard1989}. As depicted in Fig.~\ref{fig7}~(b), the MR curves at different temperatures do not collapse onto a single curve when plotted as a function of \textit{H}/$\rho$(\textit{H}=0), indicating a violation of Kohler’s rule. Such deviations may arise from temperature-dependent carrier densities or mobilities, multiple scattering mechanisms, or contributions from multiple electronic bands. \cite{Pradhan2024,Pradhan2025}. 
To account for the temperature-dependent change in carrier density, the MR data were further analyzed using the extended Kohler’s rule,  MR = \textit{f}(\textit{H}/$\eta_T$$\rho$(\textit{H}=0)), where $\eta_T$ represents the temperature-dependent  change in carrier-density \cite{Xu2021,Sankar2018}. The corresponding extended Kohler plot, shown in Fig.~\ref{fig7}~(c), demonstrates that the MR curves converge onto a single universal curve, indicating that the transport response can be scaled by incorporating $\eta_T$. The temperature dependence of $\eta_T$ is shown in the inset of Fig.~\ref{fig7}~(c).
The field dependence of the MR for $H \parallel [010]$ was also fitted using  $MR \propto H^m$, as shown in the inset of Fig.~\ref{fig7}~(a). The exponent \textit{m} increases slightly with temperature but remains close to unity over the entire temperature range up to 20 K. Within the semiclassical two-band model, \textit{m}=2 is expected for a perfectly electron-hole compensated system \cite{Fang2023}. The observed MR departs from this prediction, suggesting deviation from perfect electron-hole compensation \cite{Pradhan2024,Pradhan2025}. A similar power-law behavior is observed for $H \parallel [001]$, as illustrated in Fig.~\ref{fig7}~(d). This is further corroborated by the Hall resistivity data, where the nonlinear behaviour suggested involvement of more than one type of carriers.

\section{Conclusion}
In summary, we systematically investigated the complex magnetic behaviour and field-driven magnetic phase evolution of $\mathrm{TbTi_3Bi_4}$ by detailed temperature- and field-dependent magnetization, specific heat, and electrical transport measurements along different crystallographic directions. Our results reveal rich magnetic phases for magnetic field along the [100] direction, and demonstrate the influence of the field-induced magnetic phases on the electrical transport properties in this compound. While the \textit{dc }magnetization reproduces the previously reported sequence of field-induced metamagnetic transitions and associated fractional magnetization plateaus, our results indicate that these two transitions are fundamentally different in their dynamical behaviour. The frequency-dependent \textit{ac} susceptibility shows slow spin dynamics at the second metamagnetic transition, while no such behaviour is observed at the first metamagnetic transition, indicating distinct mechanisms underlying the rearrangement of spin structures across these two transitions. This is further confirmed by the magnetotransport measurements, where a peak-like feature in the Hall resistivity appears in the field range associated with the second metamagnetic transition. This peak-like feature may be interpreted as an enhanced anomalous Hall response arising from possible domain-wall related scattering or other non-equilibrium magnetic textures.
In contrast to the [100] direction, for field applied along the [010] and [001] directions, the magnetoresistance is positive and non-saturating. For a field along [010], it violates Kohler scaling, while improved data collapse is obtained using extended Kohler scaling with an effective temperature-dependent scaling factor. Together with the non-linear Hall response, this suggests a complex and anisotropic magneto-transport behaviour in $\mathrm{TbTi_3Bi_4}$, potentially involving multiple scattering mechanisms, temperature-dependent transport parameters, or contributions from multiple electronic bands.
These measurements demonstrate that apparently similar metamagnetic transitions can possess fundamentally different dynamical character, which can be revealed only when investigated with combined probes of magnetization, thermodynamic and electrical transport. Our findings clearly identify $\mathrm{TbTi_3Bi_4}$ as a promising centrosymmetric magnetic metal in which frustrated magnetism, metamagnetic phase transitions, spin dynamics and unconventional electronic transport are strongly intertwined. 

\ack

We acknowledge Max Planck Society for the funding support under Max Plank-India partner group project, Science and Engineering Research Board, DST, Government of India, via grant no: CRG/2022/001826 and Defence Research and Development Organisation (DRDO), Ministry of Defence, Government of India (Project no: DFTM/033203/P/41/JATC-P2QP-17). Sneh would like to thank Prime Minister Research fellowship grant (PMRF Id: 1403233) for the research
support and fellowship. Authors also thank the Central Research Facility (CRF), IIT Delhi, and the Nanoscale Research Facility at IIT Delhi for providing the materials characterization facility, like PPMS, MPMS, EDX, XRD, etc.

\data

All data that support the findings of this study are included within the article (and any supplementary files).

\bibliographystyle{iopart-num}
\bibliography{ref}

\end{document}